\documentclass[prd,aps,secnumarabic,superscriptaddress,floatfix,preprintnumbers]{revtex4-1}
\usepackage[dvips]{graphicx,color}
\usepackage{dcolumn}
\usepackage{amsmath}
\usepackage{rotating}
\def\Dot{\!\cdot\!}
\def\al{\alpha}
\def\be{\beta}
\def\ga{\gamma}
\def\de{\delta}
\def\De{\Delta}
\def\ka{\kappa}
\def\la{\lambda}

\def\ro{\rho}

\def\sig{\sigma}

\def\part{\partial}
\def\arccosh{\mbox{\rm arccosh}}

\def\Li{{\rm Li}}
\def\Re{\mbox{\rm Re}}

\newcommand{\mathsym}[1]{{}}
\newcommand{\unicode}[1]{{}}

\begin{document}

\title{Relativistic corrections to vacuum polarization contributions in muonic hydrogen}
\author{Wayne W. Repko}\email{repko@pa.msu.edu}
\affiliation{Department of Physics and Astronomy, Michigan State University, East Lansing, MI 48824}
\date{\today}
\begin{abstract}
The method proposed by Kinoshita and Nio to compute higher order vacuum polarization contributions to the Coulomb potential in muonic hydrogen is generalized to obtain relativistic corrections to their results.
\end{abstract}

\maketitle

\section{Introduction}
The usual calculations of the $e^+e^-$ vacuum polarization corrections to the Coulomb potential in muonic hydrogen are made using non-relativistic Schr\"odinger wave functions. These corrections have be calculated to orders $\al^3$ \cite{Pachucki,Pachucki_1}, $\al^4$ \cite{Pachucki,Pachucki_1,Borie_1} and $\al^5$ \cite{KN}. The experimental measurements \cite{nature,science} are sensitive to relativistic corrections to these results. This effect on the order $\al^3$ contribution has been estimated \cite{Borie_1,Borie_2} using Dirac wave functions with the Bohr radius defined by the muon-proton reduced mass and by others \cite{jentschura,kar} using extensions of non-relativistic techniques. It is of some interest to obtain a similar estimate for the $\al^4$ correction. It appears that the most convenient way to do this is to use the method introduced by Kinoshita and Nio \cite{KN}, the authors who derived the order $\al^5$ non-relativistic vacuum polarization corrections. 

In the next Section, the non-relativistic treatment is reviewed and the relativistic generalization is presented. This is followed by some concluding remarks.

\section{Relativistic corrections}
\subsection{Non-relativistic formulation}
In Ref.(\cite{KN}), the evaluation of the $\al^5$ vacuum polarization correction is formulated in momentum space. This is convenient because the various order irreducible vacuum polarization contributions $\Pi_f^{(n)}(\vec{k})$ are usually obtained from diagrammatic calculations. In the non-relativistic formulation, the vacuum polarization correction $\De E$ to the Coulomb potential is given by expansion
\begin{eqnarray} \label{DelE}
\De E &=& -\frac{e^2}{(2\pi)^3}\int\!\!d^{\,3}k \,\frac{1}{\vec{k}^{\,2}}\left(\Pi^{(2)}_f(\vec{k}^{\,2}) +\Pi^{(2)}_f(\vec{k}^{\,2})\Pi^{(2)}_f(\vec{k}^{\,2})+\Pi^{(4)}_f(\vec{k}^{\,2}) \right.\nonumber \\ 
&&\left. +\Pi^{(2)}_f(\vec{k}^{\,2})\Pi^{(2)}_f(\vec{k}^{\,2})\Pi^{(2)}_f(\vec{k}^{\,2})+ 2\Pi^{(2)}_f(\vec{k}^{\,2})\Pi^{(4)}_f(\vec{k}^{\,2})+\Pi^{(6)}_f(\vec{k}^{\,2})+ \ldots \right)\ro(\vec{k}^{\,2}a^2)\,,
\end{eqnarray}
where $a$ denotes the Bohr radius $(\mu\al)^{-1}$ and $\ro(\vec{k}^{\,2}a^2)$ is the Fourier transform of the non-relativistic probability density
\begin{equation} \label{rho}
\ro(\vec{k}^{\,2}a^2)=\int\!\!d^{\,3}r\langle |\psi(\vec{r})|^2\rangle e^{-i\vec{k}\cdot\vec{r}}\,.
\end{equation}
The angular brackets indicate an average over the degenerate $p$ states and for the $n=2$ states \cite{KN}
\begin{equation}
\ro_{2S}(\vec{k}^{\,2}a^2)=\frac{1-3\vec{k}^{\,2}a^2+2(\vec{k}^{\,2}a^2)^2}
{(1+(\vec{k}^{\,2}a^2)^2)^4}\,, \qquad \ro_{2P}(\vec{k}^{\,2}a^2) =\frac{1-(\vec{k}^{\,2}a^2)^2}{(1+(\vec{k}^{\,2}a^2)^2)^4}\,.
\end{equation}

Note that Eq.\,(\ref{DelE}) is an order by order expansion in powers of $\al$ that contains both reducible e.g. $\Pi^{(2)}_f(\vec{k}^{\,2})\Pi^{(2)}_f(\vec{k}^{\,2})$ and irreducible e.g. $\Pi^{(4)}_f(\vec{k}^{\,2})$ contributions. The sum of these contributions can be obtained if one uses the dispersion representation of K\"allen \cite{GK} and Lehmann \cite{HL}. The vacuum polarization corrections to the Coulomb potential take the form
\begin{equation} \label{corr}
D(\vec{k}^{\,2})=-\frac{e^2}{\vec{k}^{\,2}}-e^2\int_{4m_e^2}^\infty\!\!\frac{d\la}{\la} \frac{\De (\la)}{\la + \vec{k}^{\,2}}\,,
\end{equation}
where
\begin{equation}
\Delta(q^2)=\frac{(2\pi)^3}{3q^2}\sum_n\de^{(4)}(q-q_n)\langle0|j_\mu(0)|n\rangle \langle n|j^\mu(0)|0\rangle\,.
\end{equation}
For any order in $e^2$, the integral in Eq.\,(\ref{corr}) results in 
\begin{equation} \label{Pin}
-e^2\int_{4m_e^2}^\infty\!\!\frac{d\la}{\la} \frac{\De^{(n)}(\la)}{\la + \vec{k}^{\,2}}=-\frac{e^2}{\vec{k}^{\,2}}\Pi^{(n)}(\vec{k}^{\,2})\,,
\end{equation}
where $\Pi^{(n)}(\vec{k}^{\,2})$ is the sum of all $n/2$ loop vacuum polarization corrections, both reducible and irreducible. 

For $n=2$, $\De^{(2)}(\la)$ is
\begin{equation}
\De^{(2)}(\la)=\frac{\al}{3\pi}(1+2m_e^2/\la)\sqrt{1-4m_e^2/\la}\; \theta(\la-4m_e^2)\,,
\end{equation}
and, after completing the $\la$ integral, $\Pi^{(2)}(\vec{k}^{\,2})$ is the one loop correction $\Pi^{(2)}_f(\vec{k}^{\,2})$. Using the dispersion representation for $\Pi^{(2)}(\vec{k}^{\,2})/\vec{k}^{\,2}$, the expression for the leading vacuum polarization correction to the Coulomb potential can be expressed as
\begin{equation}
\De E^{(2)} = -\frac{2\mu\al^3}{3\pi^2}\int_4^\infty\!\!\frac{dx}{x}(1+2/x) \sqrt{1-4/x}\int_0^\infty\!\!\frac{dy\,y^2\left(\ro_{2P}(y)-\ro_{2S}(y)\right)} {x(\be^2x+y^2)}\,,
\end{equation}
where $\be=m_ea$. The remaining integrals can be evaluated using Mathematica's NIntegrate and give the familiar result \cite{Pachucki,Pachucki_1} 205.007 meV.

The correction of order $\al^4$ can be obtained using the expression for $\De^{(4)}(\la)$ derived by K\"allen and Sabry \cite{KS}. Explicitly, \begin{equation}
\De E^{(4)}= \frac{\mu\al^4}{\pi^3}\int_4^\infty\!\!\!dx\!\int_0^\infty\!\!\!dy\, \frac{\De^{(4)}(\sqrt{x}/2)\,y^2(\ro_{2P}(y)-\ro_{2S}(y))}{x\,(\be^2\,x+y^2)}\,,
\end{equation}
where
\begin{eqnarray}
\Delta^{(4)}(x) &=& \left(\frac{13}{54 x}+\frac{7}{108 x^3}+\frac{2}{9 x^5}\right) \sqrt{x^2-1}+\left(\frac{4}{3 x}+\frac{2}{3 x^3}\right) \sqrt{x^2-1} \log\left(8x \left(x^2-1\right)\right) \nonumber\\ [6pt]
&&+\left(-\frac{44}{9}+\frac{2}{3 x^2}+\frac{5}{4 x^4}+\frac{2}{9 x^6}\right) \arccosh(x)+\left(-\frac{8}{3}+\frac{2}{3
x^4}\right)\Bigg[\frac{2 \pi ^2}{3} \nonumber\\ [6pt]
&&-\arccosh^2(x)-\log\left(8x(x^2-1)\right)\arccosh(x)-2\Re[\Li_2\left((x+\sqrt{x^2-1})^2\right)] \nonumber \\ [2pt]
&&+\Li_2\left(-(x-\sqrt{x^2-1})^2\right)\Bigg]\,.
\end{eqnarray}
Using NIntegrate to evaluate the integrals gives the usual result \cite{Pachucki,Borie_1} $\De E^{(4)}=1.50795$ meV. The advantage here is that the bubble and the irreducible contributions need not be computed separately. 
\subsection{Relativistic corrections}
To obtain the relativistic corrections to the $\al^3$ and $\al^4$ vacuum polarization corrections in the spirit of the muonic hydrogen analysis, all that is necessary is to replace $\langle |\psi(\vec{r})|^2\rangle$ in Eq.\,(\ref{rho}) with the Dirac wave functions $\langle (\psi_\ka^{\,\mu_\ka}(\vec{r}))^\dag \psi_\ka^{\,\mu_\ka}(\vec{r})\rangle$. The general form of $\psi_\ka^{\,\mu_\ka}(\vec{r})$ is
\begin{equation}
\psi_\ka^{\,\mu_\ka}(\vec{r})=\left(\begin{array}{c}
                                    g_\ka(r)\chi_\ka^{\mu_\ka} \\ [4pt]
                                    -if_\ka(r)\sig\Dot\hat{r}
                                    \chi_\ka^{\mu_\ka}   
                                    \end{array}\right)\,, 
\end{equation}
so the expression for $\ro_\ka(\vec{k})$ is
\begin{equation}
\ro_\ka(\vec{k})=\int\!\!d^{\,3}r\left(g_\ka^2(r)+f_\ka^2(r)\right)\langle (\chi_\ka^{\mu_\ka})^\dag\,\chi_\ka^{\mu_\ka}\rangle\,e^{-i\vec{k}\cdot\vec{r}}\,.
\end{equation}
The angular integral, when averaged over $\mu_\ka$, gives $\sin(kr)/(kr)$  for all of the $n=2$ states and
\begin{equation}
\ro_\ka(k)=\frac{1}{k}\int_0^\infty\!\!dr\,r\left(g_\ka^2(r)+f_\ka^2(r)\right) \sin(kr)\,.
\end{equation}
The radial wave functions for the $n=2$, $2s_{1/2}$, $2p_{1/2}$ and $2p_{3/2}$, have $\ka=-1$, $\ka=1$ and $\ka=-2$, respectively, and can be found Rose \cite{Rose}. For example, $\ro_{-1}(k)$ is
\begin{eqnarray} \label{rhok}
\rho_{-1}(k^2) &=& 0.99999\, _2F_1\!\left[1.49997,1.99997,\frac{3}{2},-2.08229 k^2\right] -2.99996\,_2F_1\!\left[1.99997,2.49997,\frac{3}{2},-2.08229
k^2\right] \nonumber \\
&& +2.99997 _2F_1\!\left[2.49997,2.99997,\frac{3}{2},-2.08229 k^2\right]\,,
\end{eqnarray}
where $_2F_1(\al,\be;\ga ;z)$ is the hypergeometic function. The role of the Bohr radius $a$ is played by the square root of the coefficient of $k^2$ in Eq.\,(\ref{rhok}). This coefficient depends on the angular momentum of the state, resulting in two slightly different values of $a$, $a_{1/2}$=1.44301 MeV$^{-1}$ and $a_{3/2}$=1.44302 MeV$^{-1}$. The corresponding values of the parameter $\be$ are $\be_{1/2}=0.737379$ and $\be_{3/2}=0.737384$. With this in mind, the expressions for the relativistic corrections $\De E_\ka^{(2)}$ and $\De E_\ka^{(4)}$ are
\begin{eqnarray}
\De E_\ka^{(2)} &=& -\frac{2\al^2}{3\pi^2a_j}\int_4^\infty\!\!\frac{dx}{x}(1+2/x) \sqrt{1-4/x}\int_0^\infty\!\!\frac{dy\,y^2\left(\ro_\ka(y)-\ro_{-1}(y)\right)} {x(\be_j^2x+y^2)}-\De E^{(2)} \label{rel2}  \\ [4pt]
\De E_\ka^{(4)} &=&  \frac{\al^3}{\pi^3a_j}\int_4^\infty\!\!\!dx\! \int_0^\infty\!\!\!dy\, \frac{\De^{(4)}(\sqrt{x}/2)\,y^2\left(\ro_\ka(y)-\ro_{-1}(y)\right)} {x\,(\be_j^2\,x+y^2)}-\De E^{(4)}\,, \label{rel4}
\end{eqnarray}
with
\begin{eqnarray}
\rho_{-1}(x) &=& 0.99999\, _2F_1\!\left[1.49997,1.99997,\frac{3}{2},-x^2 \right] -2.99996\,_2F_1\!\left[1.99997,2.49997,\frac{3}{2},-x^2\right] \nonumber \\
&& +2.99997 _2F_1\!\left[2.49997,2.99997,\frac{3}{2},-x^2\right]\,, \\ [4pt]
\rho_1(x) &=& 9.98481\times 10^{-6}\, _2F_1\!\left[1.49997,1.99997,\frac{3}{2},-x^2\right]+9.98468\times 10^{-6}\, _2F_1\!\left[1.99997,2.49997,\frac{3}{2},-x^2\right]\nonumber \\ [4pt]
&& +0.99998\,_2F_1\!\left[2.49997,2.99997,\frac{3}{2},-x^2\right]\,, \\
\rho_{-2}(x) &=&_2F_1\!\left[2.49999,2.99999,\frac{3}{2},-x^2\right]\,.
\end{eqnarray}
The numerical integrations are evaluated using NIntegrate.

The leading order relativistic corrections from Eq.\,(\ref{rel2}) are
\begin{equation} \label{alf3}
\De E_1^{(2)} = 0.021\,{\rm meV}\,, \qquad \De E_{-2}^{(2)} = 0.026\,{\rm meV}\,,
\end{equation} \label{alf4}
which confirms the known results \cite{Borie_1}. The order $\al^4$ corrections from Eq.\,(\ref{rel4}) are
\begin{equation}
\De E_1^{(4)} = 0.00014\,{\rm meV}\,, \qquad \De E_{-2}^{(2)} = 0.00018\,{\rm meV}\,.
\end{equation}
These are obviously small but of the order that tends to be kept in calculations of contributions to the $n=2$ levels of muonic hydrogen.
\section{Conclusions}
The calculation of the relativistic corrections in momentum space is quite straightforward given the tools available to handle numerical integration and hypergeometic functions. In this calculation, the dispersion approach is particularly convenient because the explicit forms of the leading order $\Pi_f^{(2)}(\vec{k}^{\,2})$ or the next-to-leading order $\Pi_f^{(4)}(\vec{k}^{\,2})$ need not be known. The next-to-next-to-leading order is irrelevant when it comes to relativistic corrections. However, it is unfortunate that $\De^{(6)}(\la)$ is not available because it would simplify the calculation of the vacuum polarization correction at order $\al^5$ \cite{KN}.

\end{document}